\documentclass[reprint,aps,prb,twocolumn,floatfix]{revtex4-2}

\usepackage[utf8]{inputenc}
\usepackage[T1]{fontenc}
\usepackage{graphicx}
\usepackage{amsmath,amssymb}
\usepackage{hyperref}
\usepackage{xcolor}
\usepackage{physics} 
\usepackage{siunitx} 
\usepackage{comment}
\usepackage{float}
\begin{document}

\title{Breaking Bipartite and Time Reversal Symmetries by Fusing Porphine Unit in-between two Zigzag-edge Graphene Nanoribbons}

\author{R. K. Rohit$^{1}$, Jisvin Sam$^{2}$ and Sudipta Dutta$^{2}$}
\affiliation{$^{1}$Department of Physics, Sardar Vallabhbhai National Institute of Technology (SVNIT), Surat-395007, Gujarat, India. \\
$^{2}$Department of Physics, Indian Institute of Science Education and Research (IISER) Tirupati, Tirupati - 517619, Andhra Pradesh, India}

\begin{abstract}
\section*{abstract}

Hybrid structure of two zigzag-edge graphene nanoribbons with a fused porphine ring in between, results in two distinct nearly degenerate ground states: a semiconducting antiferromagnetic state and a conducting ferromagnetic state with unequal and opposite Fermi velocities of majority and minority spins, the former having slightly higher stability. Such ground states result from the broken bipartite symmetry induced by the porphine ring. The incorporation of different transition metal atoms in the porphine cavity reduces their energy difference but keeps their electronic properties mostly unchanged. The splitting of the $d$-orbitals in the distorted square-planar ligand field of porphine produces a high spin ground state that breaks the global time reversal symmetry ($\mathcal{T}$). The opposite Fermi velocities of the majority and minority spins in the ferromagnetic ground state and lower sensitivity of the conducting majority spin channel to the edge disorder, make this class of quasi-one-dimensional hybrid structures promising for dual spin-filtering device applications.

\end{abstract}

\maketitle


\section*{Introduction}

Introducing net magnetization in low-dimensional systems, especially in graphene based nanostructures, has been of paramount interest owing to their tunable device applications\cite{r1,r2,r3}. Incorporation of electronic correlations removes the Fermi instability in zigzag-edge graphene nanoribbons (ZGNRs), resulting in localization of spins along their edges. However, overall antiferromagnetic spin ordering between opposite edges with net zero magnetization indicates the robustness of the time reversal symmetry ($\mathcal{T}$) in such nanostructures\cite{r4,r5,r6,r7}. Considerable efforts have been made to break the $\mathcal{T}$ in terms of edge functionalization, proximity effect, doping and defect engineering to introduce net magnetization and consequent spin current in ZGNR-based systems\cite{r8,r9,r10,r11,r12,r13,r14,r15,r16}.

Introducing molecular units, especially organometallic units, that can interact with the edge spins can be an effective strategy to tailor the magnetic and electronic properties of ZGNRs\cite{r17,r18}. This is a type of edge functionalization with greater tunability, since the magnetic property of the overall system can be tuned by local modification in the molecular units. Recent experimental realization of porphine rings fused to the ZGNR edges has been a crucial development toward this direction\cite{r19,r20}.
Large $\pi$-conjugated macrocycles in porphine molecules can act as both electron donors and acceptors and can host numerous transition metal atoms in their central cavity. Such porphine molecules can be seamlessly integrated to the zigzag-edges of ZGNRs due to structural resemblance. Furthermore, they can introduce localized magnetic moments, spin exchange interactions between the edges and spin-dependent charge transfer, leading to complex magnetic textures at the interface\cite{r17,r18}. The experimental realization of porphine-fused GNR structures with a clear demonstration of spin-polarization and magnetic anisotropy tempts further exploration\cite{r21,r22,r23}.

In this study, we consider two ZGNRs of same or different widths and stitch them by fusing a porphine molecule between them in the unit cell, as shown in Fig.~\ref{fig:fig1}(a). These systems exhibit comparable stability for antiferromagnetic and ferromagnetic ground states with semiconducting and metallic electronic dispersions, respectively. Incorporation of transition metal atoms in the porphine cavity further reduces the energy gap between two ground states. The ferromagnetic and metallic ground state exhibits opposite group velocities for majority and minority spins at the Fermi energy. Understanding these properties can provide valuable insight for designing magnetically functional hybrid nanostructures for future spin-filtering and quantum nanodevice applications.


\section*{Computational Details}

All density functional theory (DFT) based first principle calculations in this work are carried out using the Vienna Ab initio Simulation Package (VASP)\cite{r24}. The interaction between valence electrons and ionic cores is treated using the projector augmented-wave (PAW) method\cite{r25}. For the exchange correlation functional, we employ the generalized gradient approximation (GGA) along with Perdew-Burke-Ernzerhof (PBE) exchange and correlation functionals\cite{r26}. The plane wave basis \cite{r27} set is expanded up to an energy cutoff of 500~\si{\electronvolt}, which ensures convergence of total energies and forces. Brillouin zone integrations are performed using a Monkhorst-Pack\cite{r28} $k$-point grid of 9 $\times$ 1 $\times$ 1, chosen after confirming that the total energy varies negligibly upon further refinement.

For structural relaxation, all atomic positions are optimized until the Hellmann-Feynman forces on each atom were less than \SI{1e-2}{\eV\per\angstrom}, and the total energy change between successive ionic relaxation steps is below \SI{1e-4}{\eV}. A vacuum spacing of at least \SI{15}{\angstrom} is included in the nonperiodic directions to avoid interactions between periodic images, which is essential for accurately modeling quasi one dimensional nanostructures. To accurately describe the localized d-electrons of transition metal (vanadium (V), iron (Fe), and copper (Cu)) atoms in our porphyrin-ZGNR hybrids, we employ the DFT+U formalism within GGA using the PBE functional, where U denotes the on-site Coulomb correlation\cite{r29}. This Hubbard correction follows the simplified rotationally invariant Dudarev scheme \cite{r30}, where only the effective on-site Coulomb interaction is $U_{eff}$ = U - J (with J = 0) is applied to d-orbitals, enhancing localization and magnetic moments without double counting penalties. The following Hubbard U values are obtained from literature benchmarks for bulk TM oxides, ensuring consistency with experimental magnetic properties and band gaps: U = 2.5~\si{\electronvolt} for V\cite{r31}, U = 3.5~\si{\electronvolt} for Fe\cite{r32}, and U = 4.5~\si{\electronvolt} for Cu\cite{r33}. Note that, we first perform the structural relaxation of the metal incorporated systems using DFT only for comparing the results with that of the systems without metal atoms. Then final optimization is performed with inclusion of U on DFT relaxed ground states to take into account the enhanced electronic correlation of transition metal atoms. We chose to present the results of V incorporation in the main manuscript and include the results on Fe and Cu incorporation in Supplementary Information (SI).

Spin polarized calculations are performed with both ferromagnetic and antiferromagnetic initial guesses with electronic convergence criterion set below \SI{1e-6}{\eV}. The self consistent charge density as obtained from the optimized geometry with Gaussian smearing of 0.015 eV is used for band structures and density of states (DOS) calculations. In addition, projected density of states (pDOS) and charge density analysis are carried out to interpret orbital contributions and bonding characteristics, using Vaspkit\cite{r34}.


\section*{results and discussion}

\begin{figure}[t]
\centering
\includegraphics[width=\columnwidth,
    height=9cm,
keepaspectratio]{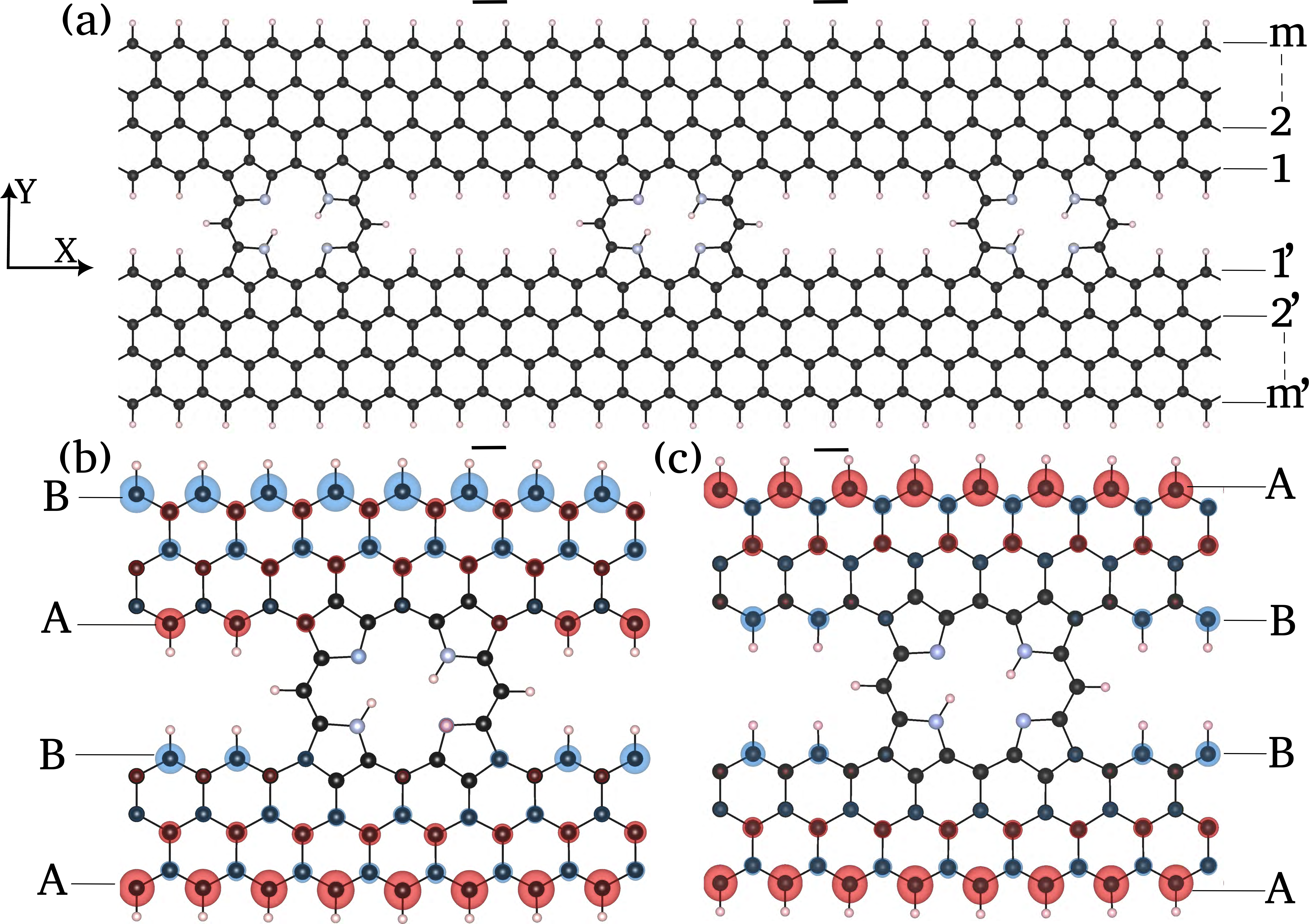}
\caption{(a) Schematic of mm'Z-Por hybrid structure with fused porphine molecules in between two ZGNR segments at the top and at the bottom, with their widths marked by the numeric values, m and m', respectively, that are the numbers of the zigzag chains along the $y$-axis. The inner edges are marked by 1 and 1' and the outer edges are marked by m and m'. The super-cell is marked with the square bracket. The ground state spin densities of (b) AFM and (c) FM ground states of 33'Z-Por hybrid show different types of spin localizations, namely majority (red) and minority (blue) spins on the two sublattices of the bipartite ZGNR segments. The AFM ground state shows two outer edges, i.e., m and m' having opposite sublattices, namely, B and A, respectively and two inner edges, i.e., 1 and 1' having opposite sublattices, namely, A and B, respectively. The outer edges (m, m') and inner edges (1, 1') of the FM ground state show same sublattices, namely A and B, respectively.}
\label{fig:fig1}
\end{figure}   

The schematic structure of the system with a fused porphine (Por) ring in between two ZGNRs is depicted in Fig.~\ref{fig:fig1}(a). We term this hybrid structure as mm'Z-Por with m and m' denoting the width of the top and bottom ZGNRs in terms of the zigzag chains in each of them, as shown in Fig.~\ref{fig:fig1}(a). The dangling bonds at both the inner edges (in 1 and 1' zigzag chains) and outer edges (in m and m' zigzag chains) of the structures are passivated by hydrogen atoms. We perform structural and electronic relaxations considering two different initial guesses: (i) A-B--A-B and (ii) A-B--B-A. Here, A and B are the two sublattice points of bipartite ZGNR lattice, that tend to localize opposite spins. In the first case, the two inner edges are considered to have opposite spins (B and A) and two outer edges are also considered to have opposite spins (A and B). In the later case, both the inner edges are considered to have same spins (B and B) and both the outer edges are considered to have same spins (A and A). Overall antiferromagnetic spin ordering in each ZGNR segment is ensured in both the initial guesses. However, after the structural and electronic relaxations, the first system stabilizes in the ground state with net zero magnetic moment, whereas the ground state of the later one shows non-zero net magnetic moment. We term these two systems as antiferromagnetic (AFM) and ferromagnetic (FM), respectively. These two ground states are energetically very close to each other and the energy difference ($\Delta$), scaled with respect to the number of atoms(N) in the super-cell reduces with increase in the width of the ZGNRs, as can be seen in TABLE~\ref{tab:Table1}. This indicates near degeneracy of the AFM and FM states for wider systems.

To gain insight about the spin ordering in both these ground states, we plot their spin densities in Fig.~\ref{fig:fig1}(b), (c), respectively. Both the AFM and FM ground states exhibit parallel spin alignment along the same edge and antiparallel spin alignment between two opposite edges of each ZGNR segment, maintaining the bipartite symmetry. However, the AFM ground state shows antiferromagnetic alignment of spins between the two inner edges and between the two outer edges (see Fig.~\ref{fig:fig1}(b)), leading to net zero magnetic moment. On the contrary, the FM ground state shows parallel spin alignment between the two inner edges and between the two outer edges (see Fig.~\ref{fig:fig1}(c)). However, higher spin moments on outer edges as compared that on inner edges induce a net magnetization of \SI{1.37}{\mu_B} in the FM ground state, thereby breaking the $\mathcal{T}$. This is in contrast with a single ZGNR where both the symmetries remain preserved with an exponential decay of spin moments from edge towards bulk. Therefore, to get further insight of this, we consider a ZGNR with 8 zigzag chains along the cross ribbon direction and scoop out the middle portion in such a way that the final structure appears similar to the 33'Z-Por structure, differing structurally in the middle region, as can be seen in Fig.~\ref{fig:S1}. In this scooped-ZGNR system, both the bipartite symmetry and $\mathcal{T}$ remains preserved, resembling the pristine ZGNR properties. This observation establishes the role of the porphine ring in breaking the bipartite symmetry between the top and bottom ZGNR segments, thereby breaking the global $\mathcal{T}$ in the FM ground state. This observation remains unchanged with increase in width of individual ZGNR segments in these hybrid structures (see Fig.~\ref{fig:S2}). Such broken bipartite symmetry due to the presence of porphine ring is responsible for negligible energy difference between the AFM and FM ground states of the hybrid structure as well.

\begin{table}[t]
\centering
\caption{Ground state energy difference ($\Delta$), ground state energy difference per atom ($\Delta/N$) corresponding to nm'Z-Por hybrids in FM and AFM states with varying n and m'.}
\label{tab:Table1}

\begin{tabular}{ccc}
\hline\hline
System & $\Delta$(m\si{\electronvolt}) = $E_{FM} - E_{AFM}$ & $\Delta/N$ \\
\hline
33'Z-Por & 39.4 & 0.30 \\
43'Z-Por & 46.4 & 0.30 \\
44'Z-Por & 43.1 & 0.26 \\
55'Z-Por & 42.7 & 0.22 \\
\hline\hline
\end{tabular}

\end{table}

\begin{figure}[t]
\centering
\includegraphics[width=\columnwidth,
    height=9cm,
keepaspectratio]{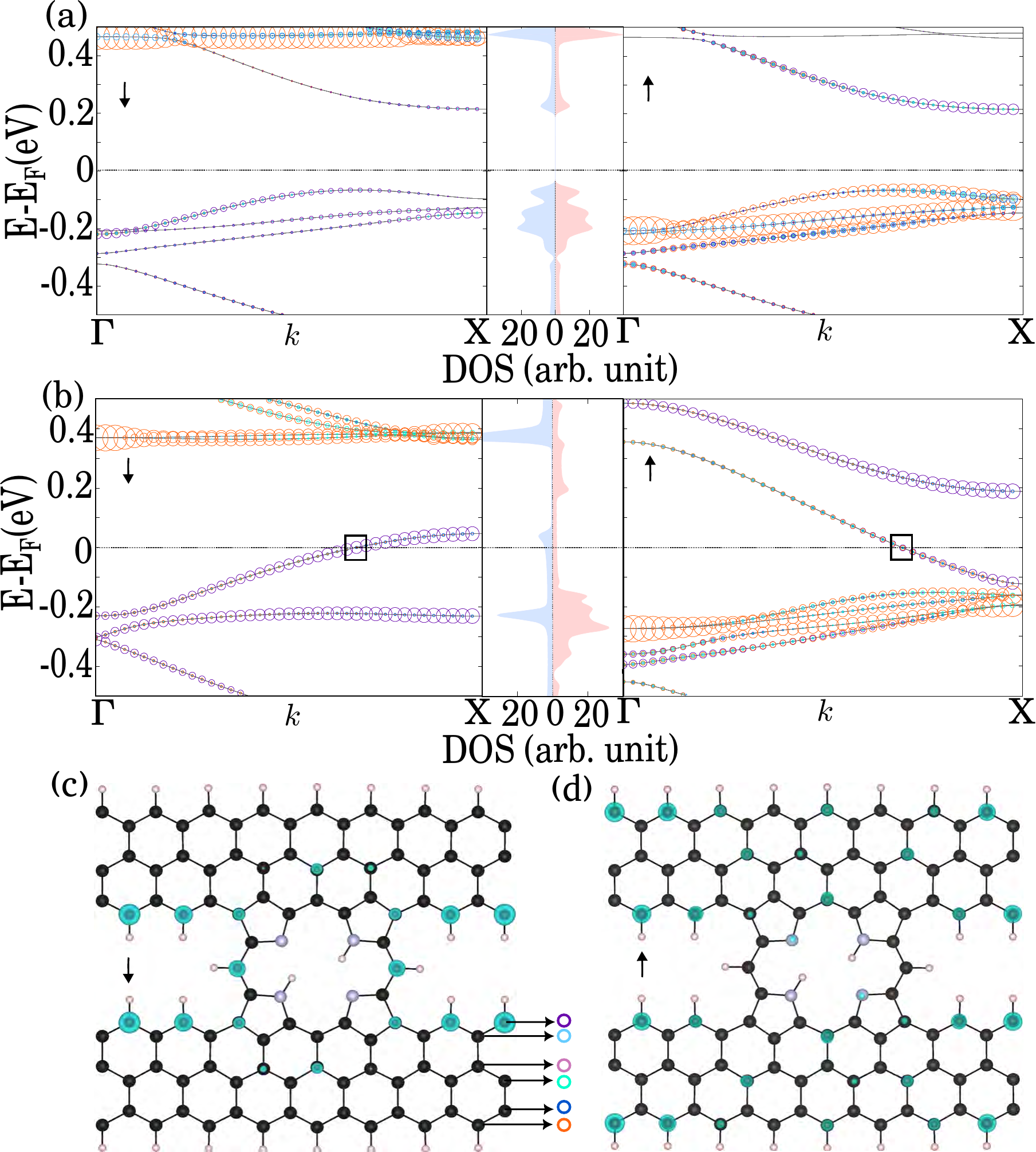}
\caption{Spin polarized $k$-resolved projected band structures for (a) AFM and (b) FM ground states of 33'Z-Por hybrid. Left and right panels show minority and majority spin band dispersions. The middle panels show spin polarized density of states (DOS) with blue and red shades representing minority and majority spins, respectively. (c) and (d) depict the probability density distributions corresponding to the states marked in boxes in the minority and majority spin band dispersions of FM state. These two states correspond to the $k$-point where the minority and majority spin bands cross the Fermi energy. Different colored circles are used to represent zigzag chain-wise normalized carbon $p_{z}$ orbital contributions in the fat bands. Those colored circles are shown in the bottom ZGNR segment in (c). The radius of the circles in the fat bands are proportional to their contributions.}
\label{fig:fig2}
\end{figure}

Further investigation of the band structure for the AFM state of 33'Z-Por hybrid shows semiconducting band gap of $\sim0.28$~\si{\electronvolt} with degenerate down and up spin bands (Fig.~\ref{fig:fig2}(a)), where the down spin bands near the Fermi energy originate from one inner and one outer edge, as observed from the fat band analysis. The up spin bands near Fermi energy too have contributions from the other set of inner and outer edges. Gradual increase in the widths of the ZGNR segments in either sides of porphine unit does not alter the band dispersions or the band gap significantly (see Fig.~\ref{fig:S3}).

On the other-hand, 33'Z-Por hybrid in FM state shows asymmetric dispersions and DOS for minority and majority spins due to $\mathcal{T}$ breaking [see Fig.~\ref{fig:fig2}(b)]. Here both the spin channels are conducting with Fermi velocities of $2.6 \times 10^{5}$ and $-5.3 \times 10^{5}$ $ms^{-1}$, respectively (see TABLE~\ref{tab:Table2} in SI). It is to be noted that the minority and majority spins are exhibiting opposite group velocities of conducting electrons and therefore can lead to dual spin filtering based on the polarity of the applied bias. As can be seen from the fat band analysis in Fig.~\ref{fig:fig2}(b), the minority spin band that crosses the Fermi energy in case of 33'Z–Por is predominantly contributed from the inner edges. On the contrary, the majority spin band that crosses the Fermi energy has contribution from both the inner and outer edges and also from the bulk atoms. Therefore, the conducting majority spins are delocalized essentially over the ZGNR segments. For further verification of the origin of the conducting spins, we plot the square of the wave functions at the $k$-points (marked by boxes in Fig.~\ref{fig:fig2}(b)) where the minority and majority spin bands cross the Fermi energy in Fig.~\ref{fig:fig2}(c) and (d), respectively. These further confirm the localization of minority spins on the inner edges and delocalization of majority spins on the ZGNR segments.

The topological edge states, the characteristic of pristine ZGNRs\cite{r35, r36, r37, r38}, appears $\sim 0.3$~\si{\electronvolt} below the Fermi energy, as can be seen in Fig.~\ref{fig:fig2}(b). This behavior can be understood as band reordering originating from exchange splitting of the edge localized states. Within a mean field description, the interaction term produces an effective spin dependent potential. Since edge states are strongly localized at the magnetic edges, they experience a significantly larger exchange shift than bulk states. In the present system, the imbalance between spin polarized outer and inner edges produces a nonzero net exchange field, consistent with Lieb’s theorem for bipartite lattices\cite{r39}, which pushes the majority spin edge states to lower energies relative to the mixed edge-bulk hybridized bands. The low energy electronic dispersions, as shown in Fig.~\ref{fig:fig2} are contributed from the carbon $p_{z}$ electrons (see Fig.~\ref{fig:S4}).

With gradual increase in the widths of the ZGNR segments on either sides of the porphine unit, the minority spin band that crosses the Fermi energy becomes less dispersive, thereby reducing the corresponding Fermi velocity (see Fig.~\ref{fig:S5}). However, the majority spin band across the Fermi energy remains dispersive with almost unaltered Fermi velocity. The gradual localization of minority spins with increase in width leads to half-metallic behavior with open majority spin channel. Hence, the dual spin filtering effect becomes compensated as the width of the proposed hybrids increases, while the single spin filtering effect remains robust. Such transport is not expected to get affected due to edge deformities or impurities, since the conduction is not restricted through the edge states only, unlike the pristine ZGNRs.

We further investigate the hybrid system with asymmetric ZGNR segments, i.e., with different widths in either sides of the porphine ring. Our observation of conducting spin channels and non-zero magnetization remains same for the FM ground state of 43'Z-Por system Fig.~\ref{fig:S6}(a). However, the AFM ground state shows non-degenerate down and up spin dispersions, despite having net zero magnetization Fig.~\ref{fig:S6}(c). This can be attributed to the asymmetric width of the ZGNR segments and their asymmetric spin distributions. This is further confirmed from the non-degenerate down and up spin dispersions of the asymmetric scooped-ZGNR Fig.~\ref{fig:S6}(e).

\begin{figure*}
\centering
\includegraphics[width=\textwidth,
    height=20cm,
keepaspectratio]{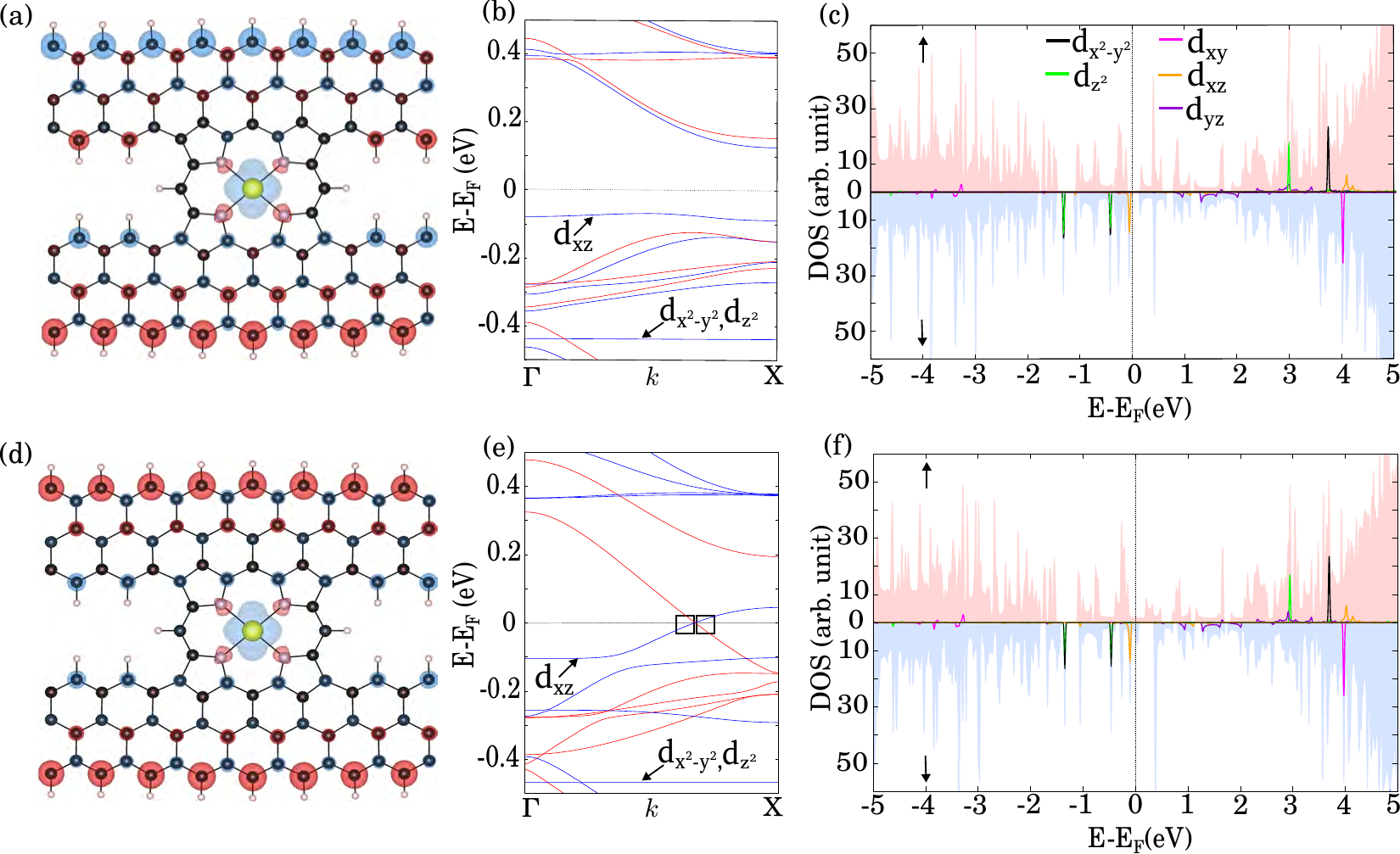}
\caption{Spin densities of the 33'Z-(V)Por hybrid in (a) AFM and (d) FM ground states. The majority and minority spins are depicted in red and blue colors, respectively with V atom localizing the minority spin. The corresponding spin resolved band structures for these states are shown in (b) and (e), respectively with same color code. The bands orginating from the V $d$-orbitals are marked with arrows. Two boxes in (e) indicate the location of the $k$-points for which the wave-functions of the majority and minority spin bands at the Fermi energy are presented in Fig.~\ref{fig:S9}(a) and (b), respectively. (c) and (f) present the total spin polarized DOS of 33'Z-(V)Por hybrid along with the individual V $d$-orbital pDOS in its AFM and FM ground states, respectively.}
\label{fig:fig3}
\end{figure*}

To influence the spin-ordering in both the AFM and FM ground states of 33'Z-Por hybrid, We further introduce the transition metal (TM) atoms in the porphine cavity. We incorporate V, Fe and Cu atoms into the pristine 33'Z-Por and model these TM atoms with on-site Coulomb correlations. We term these as 33'Z-(TM)Por in either AFM or FM state. We present the spin density, pDOS and band dispersions of 33'Z-(V)Por in AFM and FM states in Fig.~\ref{fig:fig3} and provide the same results for 33'Z-(Fe)Por and 33'Z-(Cu)Por in Fig.~\ref{fig:S7} and Fig.~\ref{fig:S8}, respectively.

The 33'Z-(V)Por in AFM and FM states show net moments of -2.527$\mu_\mathrm{B}$ and -2.519$\mu_\mathrm{B}$, respectively on V atom (see Figs.~\ref{fig:fig3}(a) and (d)) \cite{r40, r41, r42, r43}. However, the spin densities in the ZGNR segments remain unchanged with respect to the 33'Z-Por systems in their respective magnetic ground states. Note that, for consistent representation, we chose to term the majority and minority spins as described in case of 33'Z-Por, although the TM atoms prefer to localize the minority spin moment. The incorporation of V atom reduces the energy difference between the FM and AFM super-cells to 19.2 m\si{\electronvolt} which is lower as compared to that of 33'Z-Por (shown in TABLE~\ref{tab:Table3}). This is consistent for other TMs as well (see TABLE~\ref{tab:Table3} in SI). This suggests that the incorporated metal atom in the porphine cavity favors the ferromagnetic coupling between the two inner edges and thereby providing higher stability to the A-B--B-A arrangement of two ZGNR segments. This trend indicates that the nm'Z-(TM)Por systems with wider ZGNR segments will show near degenerate ground states and it necessitates the exploration of both the AFM and FM ground states. 

The appearance of net magnetization in AFM 33'Z-(V)Por ground state breaks the $\mathcal{T}$ and lead to non-degenerate majority and minority spin bands, as shown in Fig.~\ref{fig:fig3}(b). However, the system remains as semiconducting with majority spin gap of 0.278~\si{\electronvolt} and minority spin gap of 0.195~\si{\electronvolt}. On the contrary, the FM 33'Z-(V)Por ground state shows the band dispersions similar to the 33'Z-Por system, i.e., metallic behavior for both the majority and minority spins with opposite group velocities at Fermi energy (see Fig.~\ref{fig:fig3}(e)), with comparable Fermi velocities as observed in case of 33'Z-Por system. The wave-function analyses for both the spin bands near the Fermi energy (marked by boxes in Fig.~\ref{fig:fig3}(e)) too show localization of minority spins on the inner edges and delocalization of majority spins on the ZGNR segments (see Fig.~\ref{fig:S9}). Therefore this system is expected to exhibit dual spin filtering behavior, similar to the 33'Z-Por system.

Due to the geometry of the isolated porphine ligand, the $d$-orbitals of the TM atoms in the porphine cavity experience square planar electrostatic repulsion due to four N atoms, resulting in the energy order of $d_{x^2 - y^2}$ < $d_{xz}$ $\sim$ $d_{yz}$ < $d_{z^2}$ < $d_{xy}$. The degeneracy between the $d_{xz}$ and $d_{yz}$ orbitals is expected to get lifted due the attachment of ZGNR segments attached on top and bottom (along the $y$-axis) of porphine ring \cite{r44}, with down-shift of $d_{xz}$. The relative magnitudes of the ligand field splitting energy ($\Delta^{'}$) and the intra orbital electron pairing energy (P) determine the energy splitting of $d$-orbitals and resulting spin configurations of a TM atom. Due to low ligand field of N atoms, the $d$-orbital energy splitting is expected to be lower and the 33'Z-(TM)Por systems are expected to exhibit high-spin ground state with maximal unpaired electrons. 

To investigate the same, we present the pDOS arising from the $d$-orbitals of the V atom in AFM and FM ground states of 33'Z-(V)Por system, along with the total DOS over a large energy window in Figs.~\ref{fig:fig3}(c) and (f), respectively. Although the total DOS of AFM and FM states vary significantly owing to their distinct band dispersions, the pDOS for the $d$-orbitals in these two states show almost identical appearances. We find the single occupancy of the following V $d$-orbitals in the increasing energy order: $d_{x^2 - y^2}$ < $d_{xz}$ < $d_{z^2}$ (see Fig.~\ref{fig:S10}) . The down-shift of $d_{z^2}$ even below the $d_{yz}$ is noteworthy. As can be seen, the pDOS contribution of the $d_{yz}$ orbital shows broad distribution in the conduction band, indicating significant overlap with the C $p_{z}$ states. Such overlap pushes the $d_{yz}$ orbital higher in energy and consequently making it unoccupied. The $d_{xz}$ orbital that shows minority spin pDOS contribution around $\sim$ -0.1~\si{\electronvolt}, gets hybridized with the C $p_{z}$ states of the inner edges, as evident from the mixing of $d_{xz}$ minority spin flat band and the dispersive conducting band arising from the inner edges (see Fig.~\ref{fig:fig3}(e)). This sets up ferromagnetic coupling between the two inner edges, mediated through the V atom and provides higher stabilization to the FM ground state. This can be further established in terms of isolated dispersion less V $d_{xz}$ band in AFM state around $\sim$ -0.1~\si{\electronvolt}. On the contrary, the $d_{x^2 - y^2}$ and $d_{z^2}$ orbitals in both the AFM and FM ground states gives rise to localized minority spin pDOS around $\sim$ -0.5~\si{\electronvolt}. The corresponding flat band do not get mixed with any other bands, indicating negligible interaction with any other orbitals, as can be seen in Fig.~\ref{fig:fig3}(e). The axial nature of the $d_{x^2 - y^2}$ orbital anyway ensures lowest metal-ligand interaction. The negligible interaction of $d_{z^2}$ orbital, probably due to the elongation of porphine unit along the $x$-axis energetically down-shifts this orbital below $d_{yz}$ orbital.

In case of Fe ($d^{6}$) and Cu ($d^{9}$) incorporations, we find the following energy orderings: $d_{x^2 - y^2}^{2}$ < $d_{xz}^{2}$ < $d_{z^2}^{1}$ < $d_{yz}^{1}$ and $d_{x^2 - y^2}^{2}$ < $d_{xz}^{2}$ < $d_{z^2}^{2}$ < $d_{yz}^{2}$ < $d_{xy}^{1}$, respectively. The electronic bands originating from the $d_{xz}$ orbital in both these systems appear away from Fermi energy and consequently their hybridization with the inner edge C $p_{z}$ orbitals provide higher stability to the respective FM ground states, but not to the extent of V incorporated system. Similar to the V incorporation, the $d_{yz}$ orbital shows broad pDOS in case of Fe and Cu incorporations also, a signature of the higher hybridization with the C $p_{z}$ orbitals \cite{r45, r46}. The semiconducting behavior of the AFM ground state and the conducting majority and minority spin dispersions with opposite Fermi velocities remain intact in case of Fe or Cu incorporations (see Fig.~\ref{fig:S7} and Fig.~\ref{fig:S8}). 


\section*{conclusion}

Based on first-principles calculations, we demonstrate that the porphine ring fused between two ZGNR segments of the same or different widths significantly modifies their electronic properties mainly due to the broken bipartite symmetry. It produces nearly degenerate AFM and FM ground states with distinct electronic, magnetic and transport properties. The semiconducting AFM ground state shows antiparallel spin alignments between two outer edges and between two inner edges, resulting in net zero magnetization. However, the FM ground state exhibits net magnetization and consequent broken $\mathcal{T}$ and shows parallel spin alignments between two outer edges and between two inner edges. Moreover, this FM ground state shows conducting majority and minority spin channels with opposite Fermi velocities that can be exploited for dual spin filtering applications with tuning applied bias polarity. The incorporation of TM atoms in the porphine cavity provides further stability to the FM ground state due to hybridization of TM $d_{xz}$ orbital with the inner edge C $p_{z}$ orbitals, establishing an enhanced ferromagnetic coupling between the inner edges. Further analysis of the $d$-orbital splitting in the asymmetric square-planar ligand field of the porphine moiety reveals considerable modifications of the energy ordering of the $d$-orbitals based on their high-spin electron occupancy and hybridization of frontier $d$-orbitals with the C $p_{z}$ orbitals. The resulting band dispersions still favor the dual spin filtering behavior of the FM ground state. An increase in the width of the ZGNR segments leads to half-metallic behavior with conducting spin channels spreading all over the ZGNR segments, making such electronic behavior resistant to edge disorder. Our study paves a new avenue to tune the electronic properties of graphene nanoribbons by preparing experimentally feasible hybrid structures with porphine rings and further tuning of the electronic, magnetic and transport properties based on the incorporation of varying TM atoms. This may find applications in advanced spintronics and quantum logic gate devices.

\begin{acknowledgments}
JS thanks Ministry of Education, Govt. of India for PMRF grant (No. 901571). SD acknowledges IISER Tirupati for Intramural Funding and SERB, Department of Science and Technology (DST), Govt. of India for research grant CRG/2021/001731. RKR, JS and SD acknowledge National Supercomputing Mission (NSM) for providing computing resources of ‘PARAM Brahma’ at IISER Pune, which is implemented by C-DAC and supported by the Ministry of Electronics and Information Technology (MeitY) and DST, Govt. of India.
\end{acknowledgments}

\bibliographystyle{apsrev4-2}
\bibliography{references1}

@article{r1,
  author  = {Enoki, Toshiaki and Takai, Kazuyuki},
  title   = {The edge state of nanographene and the magnetism of the edge-state spins},
  journal = {Solid State Commun.},
  volume  = {149},
  number  = {27},
  pages   = {1144--1150},
  year    = {2009}
}

@article{r2,
author = {Toshiaki Enoki and Yousuke Kobayashi and Ken-Ichi Fukui},
title = {Electronic structures of graphene edges and nanographene},
journal = {International Reviews in Physical Chemistry},
volume = {26},
number = {4},
pages = {609--645},
year = {2007},
}

@article{r3,
  author = {Joly, V. L. Joseph and Kiguchi, Manabu and Hao, Si-Jia and Takai, Kazuyuki and Enoki, Toshiaki and Sumii, Ryohei and Amemiya, Kenta and Muramatsu, Hiroyuki and Hayashi, Takuya and Kim, Yoong Ahm and Endo, Morinobu and Campos-Delgado, Jessica and López-Urías, Florentino and Botello-Méndez, Andrés and Terrones, Humberto and Terrones, Mauricio and Dresselhaus, Mildred S.},
  title = {Observation of magnetic edge state in graphene nanoribbons},
  journal = {Phys. Rev. B},
  volume = {81},
  number = {24},
  pages = {245428},
  year = {2010}
}

@article{r4,
  author  = {Nikolai A. Poklonski and Eugene F. Kislyakov and Sergey A. Vyrko and Oleg N. Bubel and Sergey V. Ratkevich},
  title   = {Electronic band structure and magnetic states of zigzag graphene nanoribbons: quantum chemical calculations},
  journal = {J. Nanophotonics},
  volume  = {6},
  number  = {1},
  pages   = {061712},
  year    = {2012}
}

@article{r5,
  author  = {Asano, Taizo and Nakamura, Jun},
  title   = {Edge-State-Induced Stacking of Zigzag Graphene Nanoribbons},
  journal = {ACS Omega},
  volume  = {4},
  number  = {26},
  pages   = {22035--22040},
  year    = {2019}
}

@article{r6,
  author  = {Hsieh, Kimberly and Kochat, Vidya and Biswas, Tathagata and Tiwary, Chandra Sekhar and Mishra, Abhishek and Ramalingam, Gopalakrishnan and Jayaraman, Aditya and Chattopadhyay, Kamanio and Raghavan, Srinivasan and Jain, Manish and Ghosh, Arindam},
  title   = {Spontaneous Time-Reversal Symmetry Breaking at Individual Grain Boundaries in Graphene},
  journal = {Phys. Rev. Lett.},
  volume  = {126},
  number  = {20},
  pages   = {206803},
  year    = {2021}
}

@article{r7,
  author  = {Moles, Pablo and Santos, Hernán and Domínguez-Adame, Francisco and Chico, Leonor},
  title   = {Tuning magnetism in graphene nanoribbons via strain and adatoms},
  journal = {Phys. Rev. Res.},
  volume  = {7},
  number  = {3},
  pages   = {033255},
  year    = {2025}
}

@article{r8,
  author  = {Jiang, Peng and Tao, Xixi and Hao, Hua and Liu, Yushen and Zheng, Xiaohong and Zeng, Zhi},
  title   = {Two-dimensional centrosymmetrical antiferromagnets for spin photogalvanic devices},
  journal = {npj Quantum Inf.},
  volume  = {7},
  number  = {1},
  pages   = {21},
  year    = {2021}
}

@article{r9,
  author  = {Yazyev, Oleg V. and Katsnelson, M. I.},
  title   = {Magnetic Correlations at Graphene Edges: Basis for Novel Spintronics Devices},
  journal = {Phys. Rev. Lett.},
  volume  = {100},
  number  = {4},
  pages   = {047209},
  year    = {2008}
}

@article{r10,
  author  = {Kumar, Vipin and Kolli, Venkata and Shukla, Shobha and Saxena, Sumit},
  title   = {Spin filtering in oxidized zigzag graphene nanoribbons},
  journal = {Diamond Relat. Mater.},
  volume  = {102},
  pages   = {107662},
  year    = {2020}
}

@article{r11,
  author = {Milivojević, Marko and Mnich, Juraj and Jureczko, Paulina and Kurpas, Marcin and Gmitra, Martin},
  title = {Ferroelectric switching control of spin current in graphene proximitized by In2Se3},
  journal = {Mater. Futures},
  volume = {5},
  number = {1},
  pages = {015201},
  year = {2026}
}

@article{r12,
  author  = {Kohda, Makoto and Okayasu, Takanori and Nitta, Junsaku},
  title   = {Spin-momentum locked spin manipulation in a two-dimensional Rashba system},
  journal = {Sci. Rep.},
  volume  = {9},
  number  = {1},
  pages   = {1909},
  year    = {2019}
}

@article{r13,
  author  = {Ozaki, Taisuke and Nishio, Kengo and Weng, Hongming and Kino, Hiori},
  title   = {Dual spin filter effect in a zigzag graphene nanoribbon},
  journal = {Phys. Rev. B},
  volume  = {81},
  number  = {7},
  pages   = {075422},
  year    = {2010}
}

@article{r14,
  title = {Intrinsic Half-Metallicity in Modified Graphene Nanoribbons},
  author = {Dutta, Sudipta and Manna, Arun K. and Pati, Swapan K.},
  journal = {Phys. Rev. Lett.},
  pages = {096601},
  year = {2009},
}

@article{r15,
  author  = {Sudipta Dutta and Swapan K. Pati},
  title   = {Half-Metallicity in Undoped and Boron Doped Graphene Nanoribbons in the Presence of Semilocal Exchange-Correlation Interactions},
  journal = {J. Phys. Chem. B},
  year    = {2008},
  volume  = {112},
  number  = {5},
  pages   = {1333--1335},
}

@Article{r16,
author ="Dutta, Sudipta and Pati, Swapan K.",
title  ="Novel properties of graphene nanoribbons: a review",
journal  ="J. Mater. Chem.",
year  ="2010",
volume  ="20",
issue  ="38",
pages  ="8207-8223",
}

@article{r17,
  author  = {Kratzer, Peter and Tawfik, Sherif Abdulkader and Cui, Xiang Yuan and Stampfl, Catherine},
  title   = {Detection of adsorbed transition-metal porphyrins by spin-dependent conductance of graphene nanoribbon},
  journal = {RSC Adv.},
  volume  = {7},
  number  = {46},
  pages   = {29112--29121},
  year    = {2017}
}

@article{r18,
  author  = {Gao, Fei and Menchón, Rodrigo E. and Garcia-Lekue, Aran and Brandbyge, Mads},
  title   = {Tunable spin and conductance in porphyrin-graphene nanoribbon hybrids},
  journal = {Commun. Phys.},
  volume  = {6},
  number  = {1},
  pages   = {115},
  year    = {2023}
}

@article{r19,
  author  = {He, Yuanqin and Garnica, Manuela and Bischoff, Felix and Ducke, Jacob and Bocquet, Marie-Laure and Batzill, Matthias and Auwärter, Willi and Barth, Johannes V.},
  title   = {Fusing tetrapyrroles to graphene edges by surface assisted covalent coupling},
  journal = {Nat. Chem.},
  volume  = {9},
  number  = {1},
  pages   = {33--38},
  year    = {2017}
}

@article{r20,
  author = {Deyerling, Joel and Pörtner, Mathias and Đorđević, Luka and Riss, Alexander and Bonifazi, Davide and Auwärter, Willi},
  title = {On-Surface Synthesis of Rigid Benzenoid- and Nonbenzenoid-Coupled Porphyrin Graphene Nanoribbon Hybrids},
  journal = {J. Phys. Chem. C},
  volume = {126},
  number = {19},
  pages = {8467--8476},
  year = {2022}
}

@article{r21,
  author  = {Chen, Qiang and Lodi, Alessandro and Zhang, Heng and Gee, Alex and Wang, Hai I. and Kong, Fanmiao and Clarke, Michael and Edmondson, Matthew and Hart, Jack and O'Shea, James N. and Stawski, Wojciech and Baugh, Jonathan and Narita, Akimitsu and Saywell, Alex and Bonn, Mischa and Müllen, Klaus and Bogani, Lapo and Anderson, Harry L.},
  title   = {Porphyrin-fused graphene nanoribbons},
  journal = {Nat. Chem.},
  volume  = {16},
  number  = {7},
  pages   = {1133--1140},
  year    = {2024}
}

@article{r22,
  author  = {Mateo, Luis M. and Sun, Qiang and Eimre, Kristjan and Pignedoli, Carlo A. and Torres, Tomas and Fasel, Roman and Bottari, Giovanni},
  title   = {On-surface synthesis of singly and doubly porphyrin-capped graphene nanoribbon segments},
  journal = {Chem. Sci.},
  volume  = {12},
  number  = {1},
  pages   = {247--252},
  year    = {2021}
}

@article{r23,
  author  = {Xiang, Feifei and Gu, Yanwei and Kinikar, Amogh and Bassi, Nicolò and Ortega-Guerrero, Andres and Qiu, Zijie and Gröning, Oliver and Ruffieux, Pascal and Pignedoli, Carlo A. and Müllen, Klaus and Fasel, Roman},
  title   = {Zigzag graphene nanoribbons with periodic porphyrin edge extensions},
  journal = {Nat. Chem.},
  volume  = {17},
  number  = {9},
  pages   = {1356--1363},
  year    = {2025}
}

@article{r24,
  author  = {Kresse, G. and Furthmüller, J.},
  title   = {Efficient iterative schemes for ab initio total-energy calculations using a plane-wave basis set},
  journal = {Phys. Rev. B},
  volume  = {54},
  number  = {16},
  pages   = {11169--11186},
  year    = {1996}
}

@article{r25,
  author  = {Kresse, G. and Joubert, D.},
  title   = {From ultrasoft pseudopotentials to the projector augmented-wave method},
  journal = {Phys. Rev. B},
  volume  = {59},
  number  = {3},
  pages   = {1758--1775},
  year    = {1999}
}

@article{r26,
  author  = {Perdew, John P. and Burke, Kieron and Ernzerhof, Matthias},
  title   = {Generalized Gradient Approximation Made Simple},
  journal = {Phys. Rev. Lett.},
  volume  = {77},
  number  = {18},
  pages   = {3865--3868},
  year    = {1996}
}

@article{r27,
  author  = {Kresse, G. and Furthmüller, J.},
  title   = {Efficient iterative schemes for ab initio total-energy calculations using a plane-wave basis set},
  journal = {Phys. Rev. B},
  volume  = {54},
  number  = {16},
  pages   = {11169--11186},
  year    = {1996}
}

@article{r28,
  author  = {Monkhorst, Hendrik J. and Pack, James D.},
  title   = {Special points for Brillouin-zone integrations},
  journal = {Phys. Rev. B},
  volume  = {13},
  number  = {12},
  pages   = {5188--5192},
  year    = {1976}
}

@article{r29,
  author  = {Hubbard, J.},
  title   = {Electron correlations in narrow energy bands},
  journal = {Proc. R. Soc. London Ser. A},
  volume  = {276},
  number  = {1365},
  pages   = {238--257},
  year    = {1963}
}

@article{r30,
  author  = {Dudarev, S. L. and Botton, G. A. and Savrasov, S. Y. and Humphreys, C. J. and Sutton, A. P.},
  title   = {Electron-energy-loss spectra and the structural stability of nickel oxide: An LSDA+U study},
  journal = {Phys. Rev. B},
  volume  = {57},
  number  = {3},
  pages   = {1505--1509},
  year    = {1998}
}

@article{r31,
  author  = {Lu, Haichang and Robertson, John},
  title   = {Density Functional Theory Studies of the Metal–Insulator Transition in Vanadium Dioxide Alloys},
  journal = {Phys. Status Solidi B},
  volume  = {256},
  number  = {12},
  pages   = {1900210},
  year    = {2019}
}

@article{r32,
  author  = {Liu, Hongsheng and Di Valentin, Cristiana},
  title   = {Band Gap in Magnetite above Verwey Temperature Induced by Symmetry Breaking},
  journal = {J. Phys. Chem. C},
  volume  = {121},
  number  = {46},
  pages   = {25736--25742},
  year    = {2017}
}

@article{r33,
  author  = {Szabová, Lucie and Skála, Tomáš and Matolínová, Iva and Fabris, Stefano and Farnesi Camellone, Matteo and Matolín, Vladimír},
  title   = {Copper-ceria interaction: A combined photoemission and DFT study},
  journal = {Appl. Surf. Sci.},
  volume  = {267},
  pages   = {12--16},
  year    = {2013}
}

@article{r34,
  author  = {Vei Wang and Nan Xu and Jin-Cheng Liu and Gang Tang and Wen-Tong Geng},
  title   = {VASPKIT: A user-friendly interface facilitating high-throughput computing and analysis using VASP code},
  journal = {Comput. Phys. Commun.},
  volume  = {267},
  pages   = {108033},
  year    = {2021}
}

@article{r35,
  author  = {Katsunori Wakabayashi and Ken-ichi Sasaki and Takeshi Nakanishi and Toshiaki Enoki},
  title   = {Electronic states of graphene nanoribbons and analytical solutions},
  journal = {Sci. Technol. Adv. Mater.},
  year    = {2010},
  volume  = {11},
  pages   = {054504}
}

@article{r36,
  author  = {Ma Luo},
  title   = {Topological edge states of a graphene zigzag nanoribbon with spontaneous edge magnetism},
  journal = {Phys. Rev. B},
  year    = {2020},
  month   = {August},
  volume  = {102},
  pages   = {075421}
}

@article{r37,
  title={Peculiar Localized State at Zigzag Graphite Edge},
  author={Mitsutaka Fujita and Katsunori Wakabayashi and Kyoko Nakada and Koichi Kusakabe},
  journal={Journal of the Physical Society of Japan},
  volume={65},
  number={7},
  pages={1920-1923},
  year={1996},
}

@article{r38,
title = {Nanoscale and edge effect on electronic properties of graphene},
journal = {Solid State Communications},
volume = {152},
number = {15},
pages = {1420-1430},
year = {2012},
author = {Katsunori Wakabayashi and Sudipta Dutta},
}

@article{r39,
  author = {Lieb, Elliott H.},
  title = {Two theorems on the Hubbard model},
  journal = {Phys. Rev. Lett.},
  volume = {62},
  number = {10},
  pages = {1201--1204},
  year = {1989}
}

@article{r40,
  author = {Wierzbowska, Małgorzata and Sobolewski, Andrzej L.},
  title = {Ferrimagnetism in 2D networks of porphyrin-X and -XO (X=Sc,...,Zn) with acetylene bridges},
  journal = {J. Magn. Magn. Mater.},
  volume = {401},
  pages = {304--309},
  year = {2016}
}

@article{r41,
  author = {Prigodin, V.N. and Raju, N.P. and Pokhodnya, K.I. and Miller, J.S. and Epstein, A.J.},
  title = {Spin-Driven Resistance in Organic-Based Magnetic Semiconductor V[TCNE]x},
  journal = {Adv. Mater.},
  volume = {14},
  number = {17},
  pages = {1230--1233},
  year = {2002}
}

@article{r42,
  author = {Oliveira, Thainá Araújo and Silva, Paloma Vieira and de Vasconcelos, Fabrício Morais and Meunier, Vincent and Girão, Eduardo Costa},
  title = {Electronic and magnetic properties of porphyrin nanoribbons with chelated metals},
  journal = {Phys. Chem. Chem. Phys.},
  volume = {26},
  number = {42},
  pages = {26943--26957},
  year = {2024}
}

@article{r43,
  author = {Singh, Harish K. and Kumar, Pawan and Waghmare, Umesh V.},
  title = {Theoretical Prediction of a Stable 2D Crystal of Vanadium Porphyrin: A Half-Metallic Ferromagnet},
  journal = {J. Phys. Chem. C},
  volume = {119},
  number = {45},
  pages = {25657--25662},
  year = {2015}
}

@article{r44,
  author = {Castillo, Elisabetta and Cargnoni, Fausto and Soave, Raffaella and Trioni, Mario},
  title = {Organic Spintronics: A Theoretical Investigation of a Graphene-Porphyrin Based Nanodevice},
  journal = {Magnetochemistry},
  volume = {6},
  pages = {27},
  year = {2020}
}

@article{r45,
  author = {Liao, Meng-Sheng and Scheiner, Steve},
  title = {Electronic structure and bonding in metal porphyrins, metal=Fe, Co, Ni, Cu, Zn},
  journal = {J. Chem. Phys.},
  volume = {117},
  number = {1},
  pages = {205--219},
  year = {2002}
}

@article{r46,
  author  = {Jae-Il Lee and others},
  title   = {Electronic and magnetic properties of copper tetraazaporphyrin influenced by oxygen: A first-principles study},
  journal = {J. Korean Phys. Soc.},
  year    = {2008},
  pages   = {3662--3666}
}

\clearpage

\appendix

\setcounter{figure}{0}
\setcounter{table}{0}
\setcounter{equation}{0}

\renewcommand{\thefigure}{S\arabic{figure}}
\renewcommand{\thetable}{S\arabic{table}}
\renewcommand{\theequation}{S\arabic{equation}}

\renewcommand{\theHfigure}{S\arabic{figure}}
\renewcommand{\theHtable}{S\arabic{table}}
\renewcommand{\theHequation}{S\arabic{equation}}

\begin{titlepage}

\centering

\vspace*{3cm}

{\Huge \textbf{Supplementary Information} \par}

\vspace{1.5cm}

{\LARGE \textbf{Breaking Bipartite and Time Reversal Symmetries by Fusing Porphine Unit in-between two Zigzag-edge Graphene Nanoribbons} \par}

\vspace{2cm}

{\large
R. K. Rohit$^{1}$, Jisvin Sam$^{2}$ and Sudipta Dutta$^{2}$ \par
$^{1}$Department of Physics, Sardar Vallabhbhai National Institute of Technology (SVNIT), Surat-395007, Gujarat, India. \par
$^{2}$Department of Physics, Indian Institute of Science Education and Research (IISER) Tirupati, Tirupati - 517619, Andhra Pradesh, India.
}

\vfill

\end{titlepage}

\begin{figure*}[b]
    \centering
    \includegraphics[width=\textwidth,height=10cm,keepaspectratio]{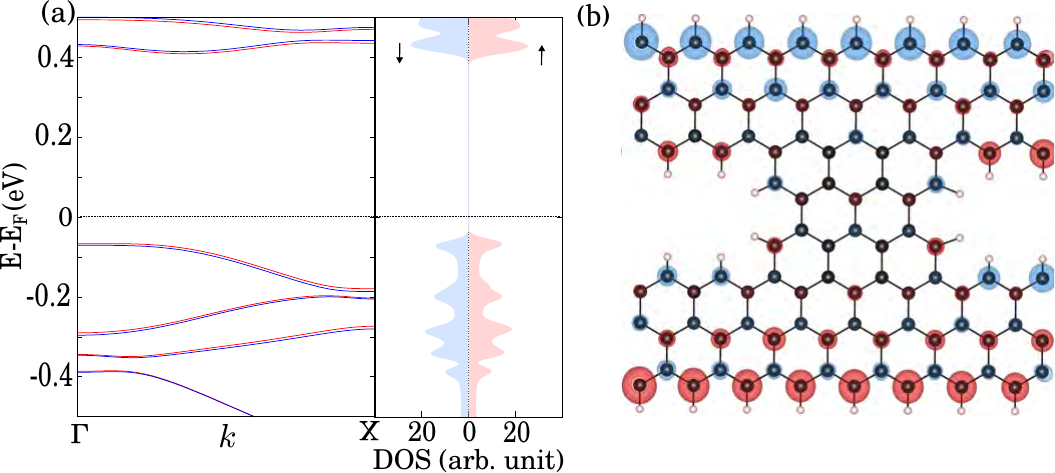}
    \caption{scooped-ZGNR (a) The spin polarized band structure and DOS scooped-ZGNR, derived from a ZGNR with 8 zigzag chains along the cross ribbon direction (b) Spin density of its ground state with red and blue colors presenting two different spins localized on A and B sublattice points of this bipartite lattice. The net magnetic moment of these system is zero.}
    \label{fig:S1}
\end{figure*}

\begin{figure*}[b]
    \centering
    \includegraphics[width=\textwidth,height=10cm,keepaspectratio]{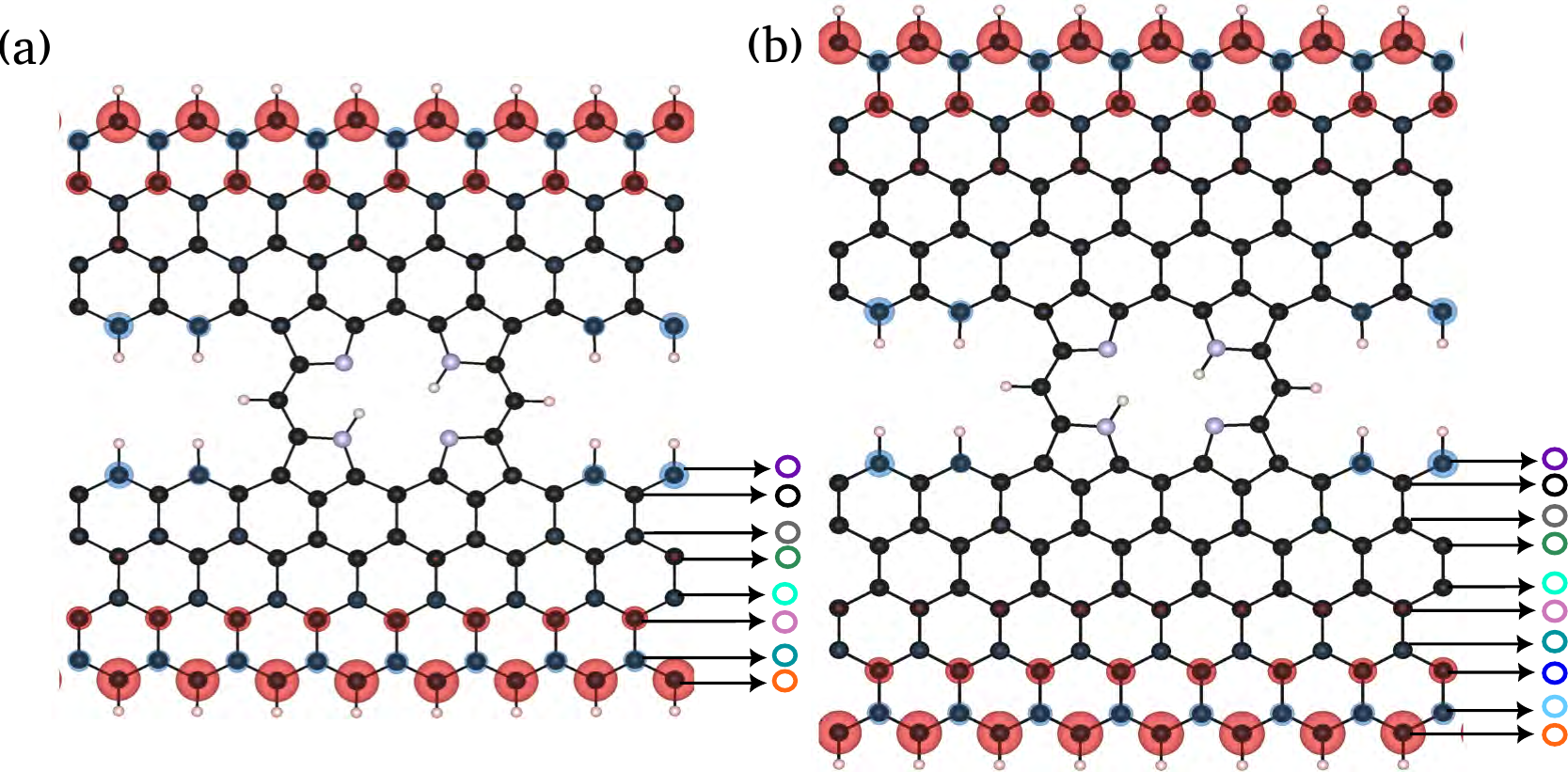}
    \caption{The spin densities of the FM ground states of (a) 44'Z-Por and (b) 55'Z- Por hybrids. Majority and minority spins are represented in red and blue colors, respectively. Two inner edges and two outer edges localize same spins.}
    \label{fig:S2}
\end{figure*}

\begin{figure*}[t]
    \centering
    \includegraphics[width=\textwidth,height=20cm,keepaspectratio]{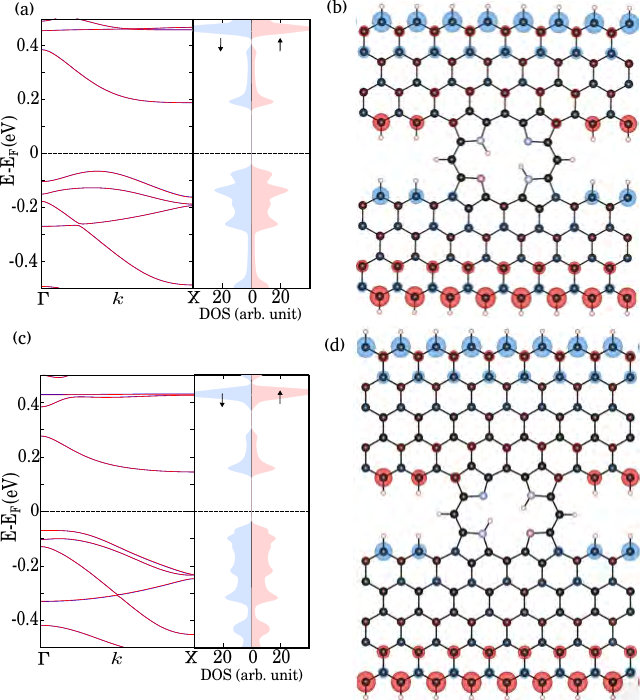}
    \caption{Spin polarized band structures and DOS of AFM ground states of (a) 44'Z-Por and (c) 55'Z-Por hybrids, where blue and red colors represent minority and majority spins, respectively. The ground state spin densities of these two systems are shown in (b) and (d), respectively. Two inner edges and two outer edges localize opposite spins.}
    \label{fig:S3}
\end{figure*}

\begin{figure*}[t]
    \centering
    \includegraphics[width=\textwidth,height=10cm,keepaspectratio]{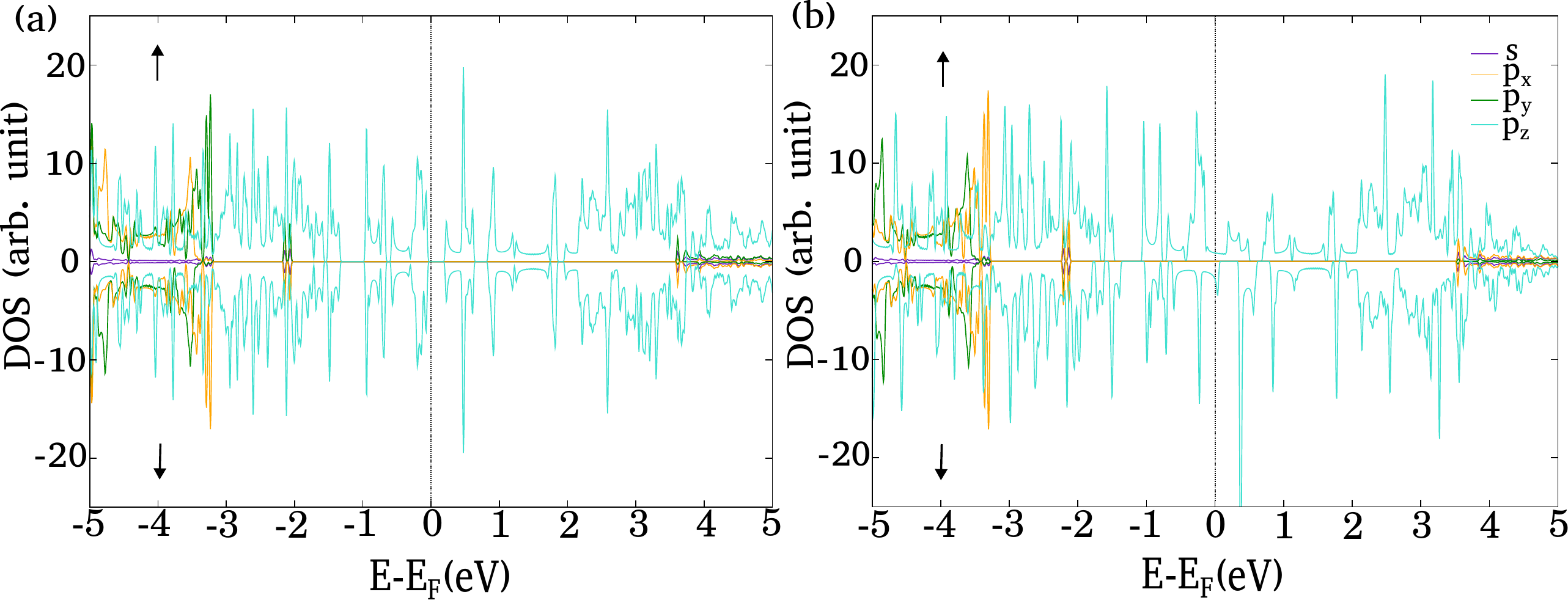}
    \caption{Spin polarized pDOS of different orbitals of all the carbon atoms (mentioned as legends) of 33'Z-Por hybrid in (a) AFM and (b) FM ground states.}
    \label{fig:S4}
\end{figure*}

\begin{figure*}[t]
    \centering
    \includegraphics[width=\textwidth,height=20cm,keepaspectratio]{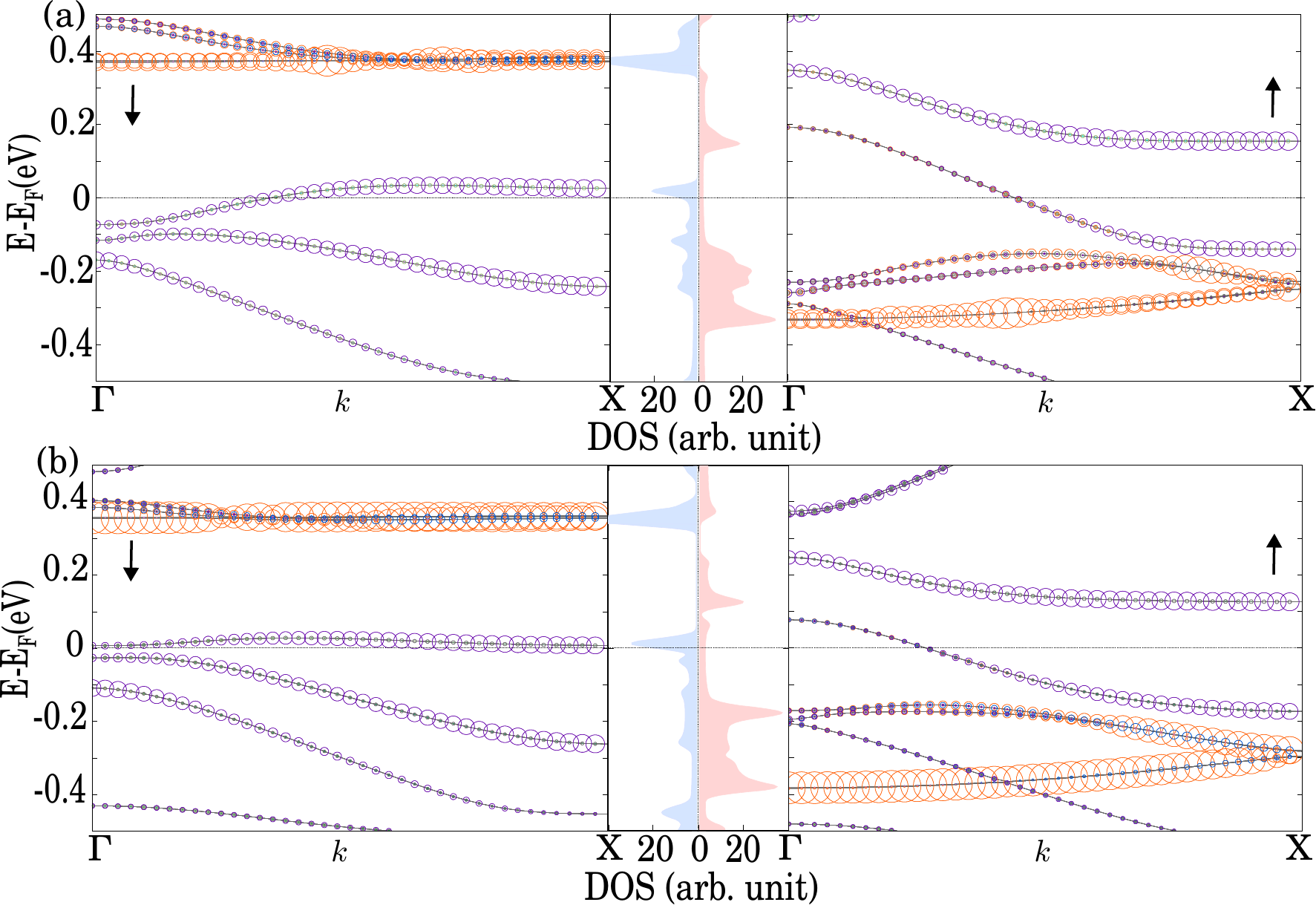}
    \caption{Spin polarized $k$-resolved projected band structure plots for (a) 44'Z-Por, (b) 55'Z-Por hybrids in FM ground states. Left and Right panels show minority and majority spin band structures. Middle panels show spin polarized DOS with blue and red shades representing minority and majority spins, respectively.  Different colored circles are used to represent zigzag chain-wise normalized carbon $p_{z}$ orbital contributions in the fat bands. Those colored circles are shown in the bottom ZGNR segment in Fig.~\ref{fig:S2}. The radius of the circles in the fat bands are proportional to their contributions.}
    \label{fig:S5}
\end{figure*}

\begin{figure*}[b]
    \centering
    \includegraphics[width=0.85\linewidth,height=0.8\textheight,keepaspectratio]{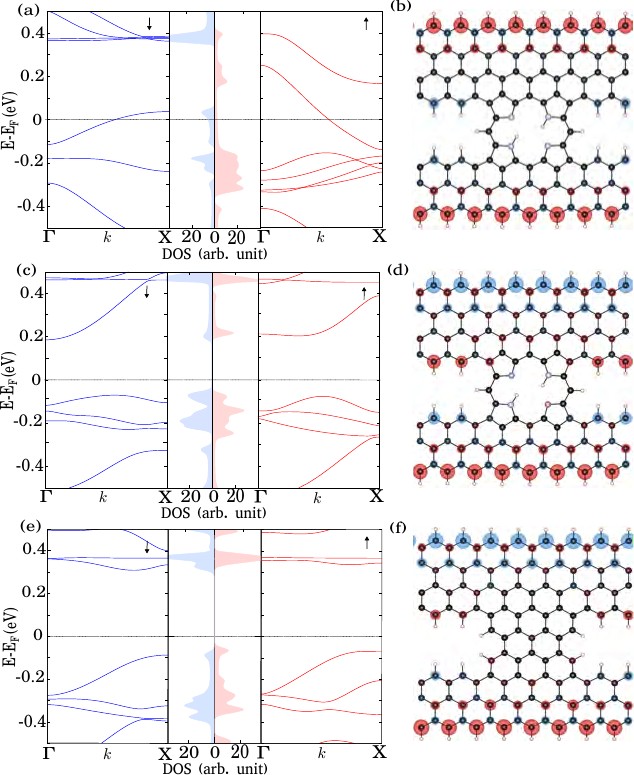}
    \caption{Spin polarized band structures and DOS for (a) FM ground state of asymmetric 43'Z-Por hybrid (c) AFM ground state of asymmetric 43'Z-Por hybrid and (e) AFM ground state of asymmetric scooped ZGNR. Their corresponding ground state spin densities are presented in (b), (d) and (f), respectively. The blue and red colors represent the minority and majority spins.}
    \label{fig:S6}
\end{figure*}

\begin{figure*}[b]
    \centering
    \includegraphics[width=\textwidth,height=25cm,keepaspectratio]{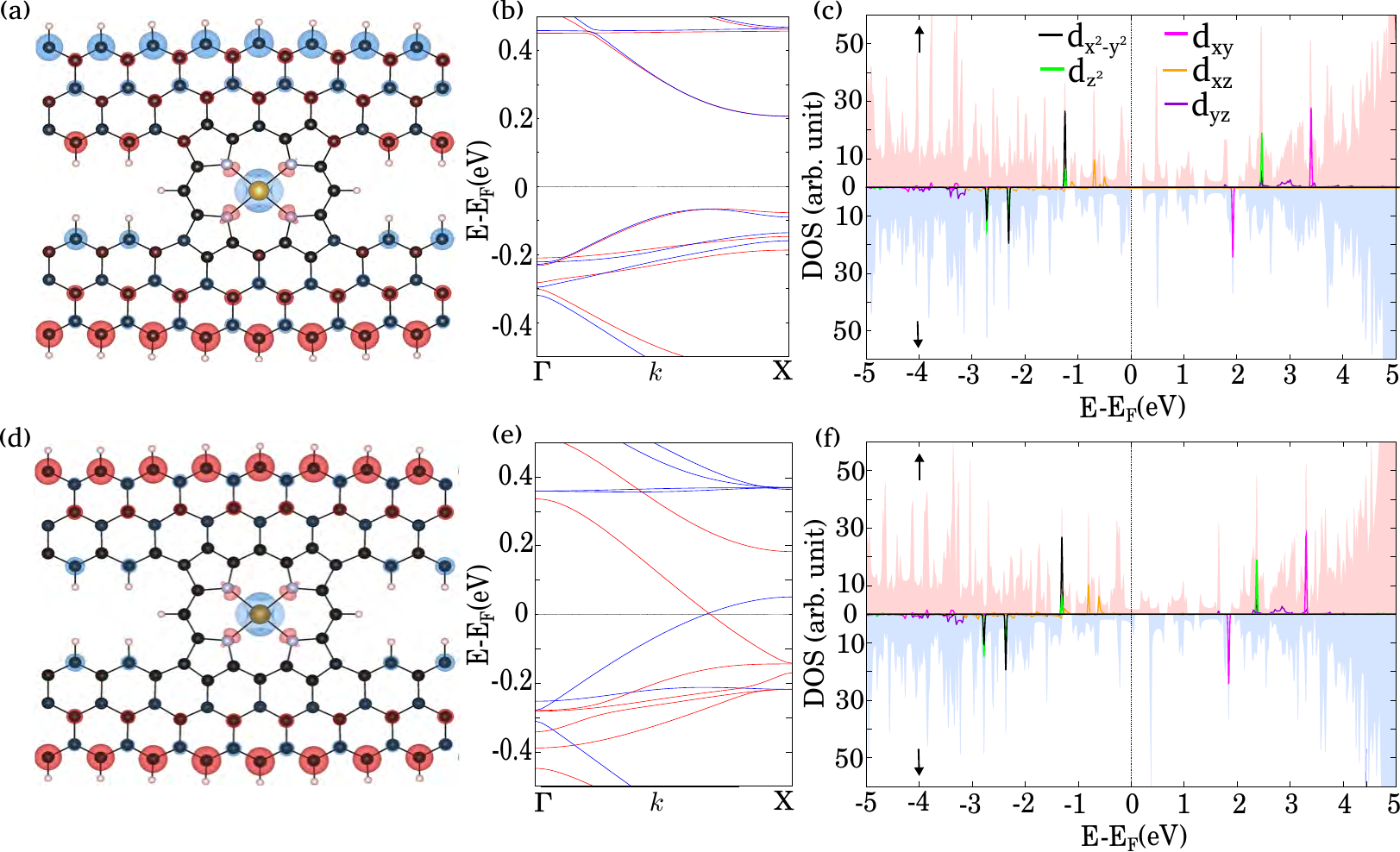}
    \caption{Spin densities of the 33'Z-(Fe)Por hybrid in (a) AFM and (d) FM ground states. The majority and minority spins are depicted in red and blue colors, respectively with Fe atom localizing the minority spin. The corresponding spin resolved band structures for these states are shown in (b) and (e), respectively with same color code. (c) and (f) present the total spin polarized DOS of 33'Z-(Fe)Por hybrid along with the individual Fe $d$-orbital pDOS in its AFM and FM ground states, respectively.}
    \label{fig:S7}
\end{figure*}

\begin{figure*}[b]
    \centering
    \includegraphics[width=\textwidth,height=25cm,keepaspectratio]{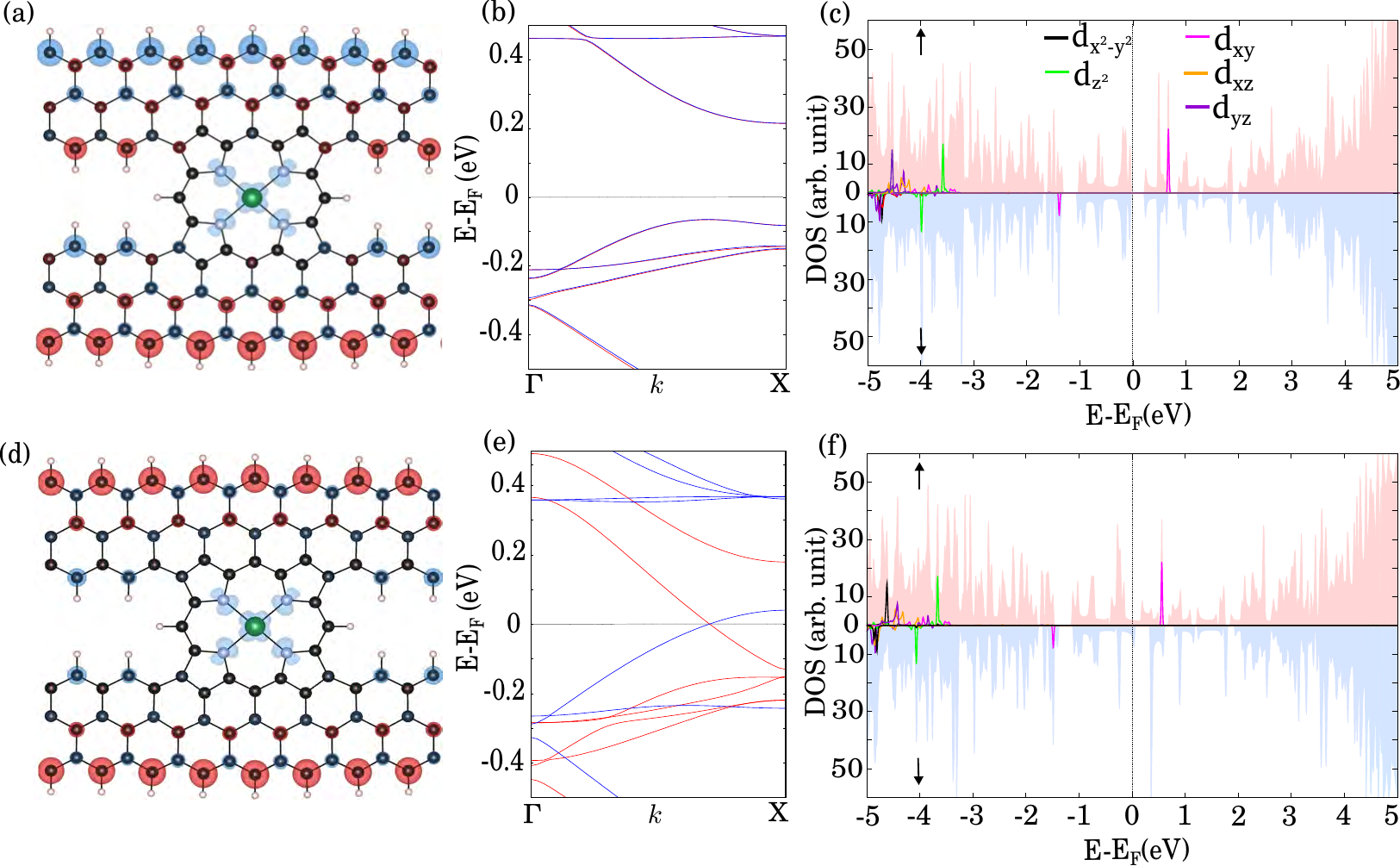}
    \caption{Spin densities of the 33'Z-(Cu)Por hybrid in (a) AFM and (d) FM ground states. The majority and minority spins are depicted in red and blue colors, respectively with Cu atom localizing the minority spin. The corresponding spin resolved band structures for these states are shown in (b) and (e), respectively with same color code. (c) and (f) present the total spin polarized DOS of 33'Z-(Cu)Por hybrid along with the individual Cu $d$-orbital pDOS in its AFM and FM ground states, respectively.}
    \label{fig:S8}
\end{figure*}

\begin{figure*}[t]
    \centering
    \includegraphics[width=\textwidth,height=10cm,keepaspectratio]{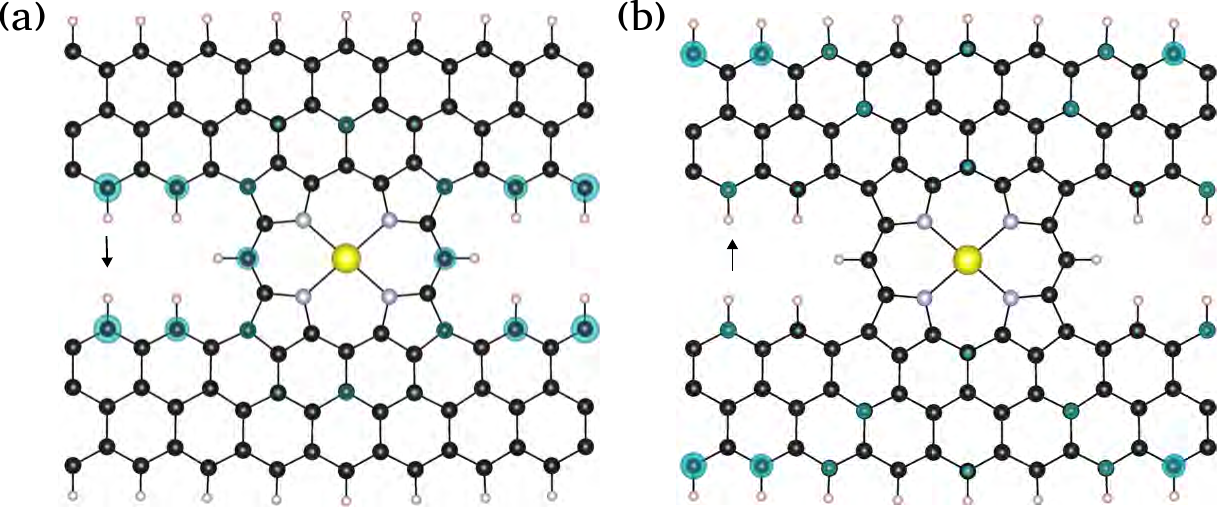}
    \caption{(a) and (b) depict the probability density distributions corresponding to the states marked in boxes in the minority and majority spin band dispersions of FM state of 33'Z-(V)Por in see Fig.~\ref{fig:fig3}(e). These two states correspond to the $k$-point where the minority and majority spin bands cross the Fermi energy.}
    \label{fig:S9}
\end{figure*}

\begin{figure*}[t]
    \centering
    \includegraphics[width=0.5\linewidth,height=0.45\textheight,keepaspectratio]{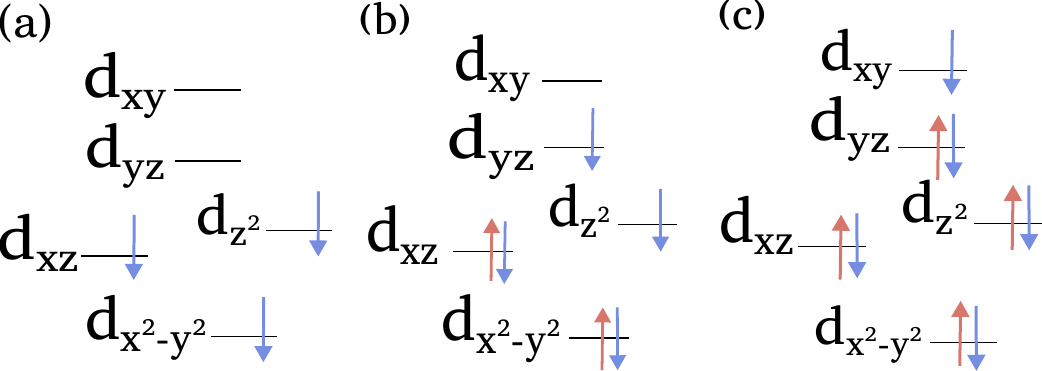}
    \caption{$d$-orbital splitting and their occupations in (a) V in 33'Z-(V)Por, (b) Fe in 33'Z-(Fe)Por and (c) Cu in 33'Z-(Cu)Por. Up arrows represnt majority spins and down arrows represent minority spins.}
    \label{fig:S10}
\end{figure*}

\clearpage
\begin{table*}[b]
\caption{Spin-resolved Fermi velocities of the 33'Z-Por and 33'Z-(TM)Por hybrids in FM ground states. }
\label{tab:Table2}
\centering
\begin{tabular}{lcc}
\hline\hline
System & $v_F^{\uparrow}$ ($10^5$ m$s^{-1}$) & $v_F^{\downarrow}$ ($10^5$ m$s^{-1}$) \\
\hline
33'Z-Por           & -5.3 & 2.6 \\
33'Z-(V)Por        & -5.4 & 2.5 \\
33'Z-(Fe)Por       & -5.2 & 2.4 \\
33'Z-(Cu)Por       & -5.4 & 2.6 \\
\hline\hline
\end{tabular}
\end{table*}

\begin{table*}[t]
\caption{Ground state energy differences($\Delta$) of 33'Z-Por and 33'Z-(TM)Por hybrids in FM and AFM states.}
\label{tab:Table3}
\centering
\begin{tabular}{lc}
\hline\hline
Hybrid & $\Delta$(m\si{\electronvolt}) = $E_{FM} - E_{AFM}$ \\
\hline
33'Z-Por           & 39.4 \\
33'Z-(V)Por        & 19.2   \\
33'Z-(Fe)Por       & 36.7 \\
33'Z-(Cu)Por       & 35.3 \\
\hline\hline
\end{tabular}
\end{table*}

\end{document}